\documentclass{llncs}

\usepackage{ucs}
\usepackage[utf8x]{inputenc}
\usepackage[T1]{fontenc}
\PrerenderUnicode{ÉéÀàÈè¼}

\usepackage{theorem}
\usepackage{amssymb}
\usepackage{xspace}

\usepackage[sectionbib]{natbib}

\usepackage{tikz}
\graphicspath{{./}{./Fig/}{./Fig_New/}}
\usepackage[center]{subfigure}
\usepackage{wrapfig}

\usepackage{h_agc}
\usepackage{h_graphics}
\usepackage{h_couleur}
\usepackage{h_math}

\newcommand{\Latin}[1]{\textsl{#1}\xspace}

\newcommand{\ie}{\Latin{i.e.}}
\newcommand{\eg}{\Latin{e.g.}}

{ \theorembodyfont{\rmfamily}
  
}

\newcommand{\Figure}[1]{Figure\,\ref{#1}}
\newcommand{\Fig}[1]{Fig.\,\ref{#1}}

\newcommand{\Appendix}[1]{Appendix\,\ref{#1}}

\newcommand{\Section}[1]{Section\,\ref{#1}}
\newcommand{\Sec}[1]{Sect.\,\ref{#1}}

\def\Pct(#1)#2{\put(#1){\makebox(0,0){#2}}}
\def\Plt(#1)#2{\put(#1){\makebox(0,0)[l]{#2}}}
\def\Pcm(#1)#2{\Pct(#1){$#2$}}

\newcommand{\DiscreteSigUn}[3]{
  \foreach \x in {0,0.25,...,#3}%
  \filldraw[fill=#1,draw=none,shift={(#2)}] (\x,\x) rectangle +(.25,.25);%
}
\newcommand{\DiscreteSigZero}[3]{
  \foreach \x in {0,0.25,...,#3}
  \filldraw[fill=#1,draw=none,shift={(#2)}] (0,\x) rectangle +(.25,.25);
}
\newcommand{\DiscreteSigMoinsUn}[3]{
  \foreach \x in {0,0.25,...,#3}
  \filldraw[fill=#1,draw=none,shift={(#2)}] (-\x,\x) rectangle +(.25,.25);
}

\newcommand{\SigAD}{\DiscreteSigUn{DarkGreen}}
\newcommand{\SigBD}{\DiscreteSigMoinsUn{Red}}
\newcommand{\SigCD}{\DiscreteSigZero{Blue}}
\newcommand{\SigDD}{\DiscreteSigZero{Brown}}
\newcommand{\SigED}{\DiscreteSigZero{Orange}}

\newcommand{\SignalContinue}[3]{%
  \draw[#1] (#2) -- (#3);
}
\def\SigAAA(#1)(#2){\SignalContinue{DarkGreen}{#1}{#2}}
\def\SigBBB(#1)(#2){\SignalContinue{Red,style}{#1}{#2}}
\def\SigCCC(#1)(#2){\SignalContinue{Blue}{#1}{#2}}
\def\SigDDD(#1)(#2){\SignalContinue{BleuFonce}{#1}{#2}}

\newcommand{\DiagETBasic}[3]{%
  \begin{scope}[x=\unitlength,y=\unitlength,shift={(#1)}]%
    \draw[->,Black] (0,0) -- (0,5.75); 
    \draw[<->,Black] (-1,0) -- (5.75,0); 
    \node[rotate=90,Black] at (-.5,2.5) {Time #3}; 
    \SigAAA(0,0)(1,1)
    \SigAAA(1,1)(2.5,2.5)
    \SigAAA(2.5,2.5)(4,4)
    \SigAAA(3,0)(4,1)
    \SigBBB(1,1)(2,0)
    \SigBBB(5,0)(4,1)
    \SigBBB(4,1)(2.5,2.5)
    \SigCCC(1,1)(1,5)
    \SigCCC(2.5,2.5)(2.5,5)
    \SigCCC(4,0)(4,1)
    \SigCCC(4,1)(4,4)
    #2
  \end{scope}
}

\usepackage{denys}

\begin{document}
%

\title{Computing in the fractal cloud:\\
modular generic solvers for\\
SAT and Q-SAT variants
}

\author{Denys Duchier
  \and
  J{\'e}r{\^o}me Durand-Lose\thanks{%
    Corresponding author \email{Jerome.Durand-Lose@univ-orleans.fr}}
  \and
  Maxime Senot}

\institute{LIFO,
 Universit\'e d'Orl\'eans, \\
B.P. 6759, F-45067 ORL\'EANS Cedex 2.
}

\maketitle

\begin{abstract}
  Abstract geometrical computation can solve hard combinatorial problems efficiently:
  we showed previously how Q-SAT can be solved in bounded space and time using
  instance-specific signal machines and fractal parallelization.  In this
  article, we propose an approach for constructing a particular \emph{generic} machine for
  the same task. This machine deploies the Map/Reduce paradigm over a fractal
  structure.  Moreover our approach is \emph{modular}: the machine is constructed
  by combining modules.  In this manner, we can easily create generic machines
  for solving satifiability variants, such as SAT, \#SAT, MAX-SAT.
\end{abstract}

\renewcommand{\abstractname}{Keywords.}
\begin{abstract}
  Abstract geometrical computation;
  Signal machine;
  Fractal;
  SAT;
  Massive parallelism;
  Model of computation.
\end{abstract}

%
\section{Introduction}
\label{sec:intro}

Since their first formulations in the seventies, problems of Boolean
satisfiability have been studied extensively in the field of computational
complexity. Indeed, the most important complexity classes can be characterized
--- in terms of reducibility and completeness --- by such problems \eg SAT for
NP \citep{cook71} and Q-SAT for PSPACE \citep{stockmeyer+meyer73}.  As such, it
is a natural challenge to consider how to solve these problems when
investigating new computing machinery (quantum, NDA, membrane, hyperbolic
spaces\dots) \citep{paun01,margenstern+morita01,alhazov+jimenez07mcu}.

This is the line of investigation that we have been following with \emph{signal
  machines} \citep{durand-lose05cie}, an abstract and geometrical model of
computation. We showed previously how such machines were able to solve SAT
\citep{duchier+durand-lose+senot10isaac} and Q-SAT
\citep{duchier+durand-lose+senot10scw} in bounded space and time.  But in both
cases, the machines were instance-specific \ie depended on the formula whose
satifiability was to be determined.  The primary contribution of the present
paper is to exhibit a particular \emph{generic} signal machine for the same task: it takes
the instance formula as an input encoded (in polynomial-time by a Turing
machine) in an initial configuration.  We further improve our previous results
by describing a \emph{modular} approach that allows us to easily construct
generic machines for other variants of SAT, such as \#SAT or MAX-SAT.


The model of signal machines, called \emph{abstract geometrical computation},
involves two types of fundamental objects: dimensionless \emph{particles} and
\emph{collision rules}.  We use here one-dimensional machines: the space is the
Euclidean real line, on which the particles move with a constant speed.
Collision rules describe what happens when several particles collide.
By representing the continuous time on a vertical axis, we obtain
two-dimensional \emph{space-time diagram}, in which the motion of the particles
are materialized by segment lines called \emph{signals}.

Signal machines can simulate Turing machines, and are thus Turing-universal
\citep{durand-lose05cie}.  They are also capable of analog computation by using
the continuity of space and time to simulate analog models such as BSS's one
\citep{durand-lose08cie,blum+shub+smale89} or computable analysis
\citep{durand-lose09uc}.  Other geometrical models of computation exist:
colored universes \citep{jacopini+sontacchi90}, geometric machines
\citep{huckenbeck89tcs}, piece-wise constant derivative systems
\citep{bournez97icalp}, optical machines \citep{naughton+woods01mcu}\dots
All these models, including signal machines, belong to a larger class of models
of computation, called \emph{unconventional}, which are more powerful than
classical ones (Turing machines, RAM, $\lambda$-calculus \ldots).  Among all
these abstract models, the model of signal machines distinguishes itself by
realistic assumptions respecting the major principles of physics --- finite
density of information, respect of causality and bounded speed of information
--- which are, in general, not respected all at the same time by other
models. Nevertheless, signal machines remain an abstract model, with no a
priori ambition to be physically realizable, and is studied for theoretical
issues of computer sciences.

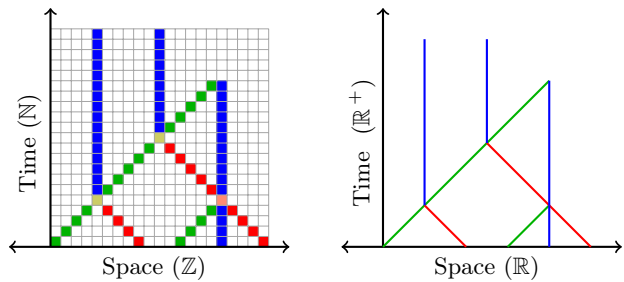
\begin{figure}[hbt]
  \centering
    \small\SetUnitlength{1.7em}
      \begin{tikzpicture}[x=\unitlength,y=\unitlength,thick]
        \SigAD{0,0}{3.75}
        \SigAD{3,0}{.75}
        \SigBD{2,0}{.75}
        \SigBD{5,0}{2.5}
        \SigDD{1,1}{0}
        \SigDD{2.5,2.5}{0}
        \SigCD{1,1.25}{3.75}
        \SigCD{2.5,2.75}{2.25}
        \SigCD{4,0}{3.75}
        \SigED{4,1}{0}
        \SigCD{4,1.25}{2.5}
        \draw[step=.25,Grey,very thin] (0,0) grid (5.25,5.25);
        \draw[->,Black] (0,0) -- (0,5.75); 
        \draw[<->,Black] (-1,0) -- (5.75,0); 
        \node[rotate=90,Black] at (-.5,2.5) {Time (\NaturalSet)}; 
        \node[Black] at (2.5,-.5) {Space (\IntegerSet)}; 
        \DiagETBasic{8,0}{%
            \node[Black] at (2.5,-.5) {Space (\RealSet)};
          }{ (\RealSet\!$^+$)} 
      \end{tikzpicture}
  \caption{From cellular automata to signal machines.}
  \label{fig:ca}
\end{figure}

As signal machines take their origins in the world of cellular automa (as
illustrated in \Fig{fig:ca}), they can also be viewed as a massively parallel
computational device. This is the approach proposed here: we put in place a
fractal compute grid, then use the Map/Reduce paradigm to distribute
the computations, then aggregate the results.

\vspace{-0.4cm}
\vspace{-0.2cm}

The Map/Reduce pattern, pioneered by Lisp, is now standard in functional
programming: a function is applied to many inputs (map), then the results are
aggregated (reduce).  Google extended this pattern to allow its distributed
computation over a grid of possibly a thousand nodes \citep{dean04}.  The idea
is to partition the input (petabytes of data) into chunks, and to process these
chunks in parallel on the available nodes.

When solving combinatorial problems, we are also faced with massive inputs;
namely, the exponential number of candidate solutions.  Our approach is to
distribute the candidates, and thus the computation, over an unbounded fractal
grid.  In this way, we adapt the map/reduce pattern for use over a grid with
fractal geometry.

Our contribution in this paper is three fold: first, we show how Q-SAT can be
solved in bounded space and time using a \emph{generic machine}, where the
input (the formula) is simply compiled into an initial configuration.  This
improves on our previous result where the machine itself depended on the
formula.  Second, we propose the first architecture for fractally distributed
computing (the \emph{fractal cloud}) and give a way to automatically shrink the data into this structure
by means of a \emph{lens device}.  Third, we show how generic machines for many
variants of SAT can be assembled by composing independent modules, which naturally
emerged from the generalization of our previous family of machines into a single machine solving Q-SAT.
 Each module can be programmed and understood independently.

The paper is structured as follow. 
Signal machines are introduced in \Section{sec:def}.
\Section{sec:fractal-cloud} presents the fractal tree structure used to achieve 
massive parallelism and how general computations can be inserted in the tree.
\Section{sec:qsat} details this implementation for a Q-SAT solver and \Section{sec:sat-variants}
explains how some variants of satisfiability problems can be solved with the
same approach.  Complexities are discussed in \Section{sec:complexities} and
conclusions and remarks are gathered in \Section{sec:conclusion}.

%
\section{Definitions}
\label{sec:def}

Signal machines are an extension of cellular automata from discrete time and
space to continuous time and space.  Dimensionless signals/particles move along
the real line and rules describe what happens when they collide.

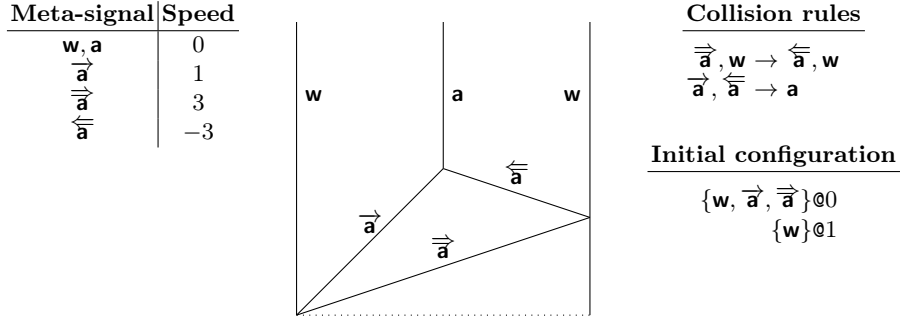
\begin{figure}[hbt]
\centering
\begin{tabular}[t]{c|c}
  \textbf{Meta-signal} & \textbf{Speed}\\\hline
  $\ST\sigwall, \ST\sigstart$ & 0\\
  $\RLO\sigstart$ & 1\\
  $\RHI\sigstart$ & 3\\
  $\LHI\sigstart$ & $-3$
\end{tabular}
\qquad
\begin{tikzpicture}[baseline=(current bounding box.north),x=2em,y=2em]
\draw[dotted] (0,0) -- (6,0);
\draw (0,0) -- (0,3) -- node[right] {$\ST\sigwall$} (0,6);
\draw (6,0) -- (6,3) -- node[left] {$\ST\sigwall$} (6,6);
\draw (0,0) -- node [above] {$\RHI\sigstart$} (6,2);
\draw (6,2) -- node [above] {$\LHI\sigstart$} (3,3);
\draw (0,0) -- node [above] {$\RLO\sigstart$} (3,3);
\draw (3,3) -- node[right] {$\ST\sigstart$} (3,6);
\end{tikzpicture}
\qquad
\begin{tabular}[t]{@{}c@{}}%
\begin{tabular}[t]{r@{ $\becomes$ }l}
\multicolumn{2}{c}{\textbf{Collision rules}}\\\hline\noalign{\vskip2mm}%
$\RHI\sigstart, \ST\sigwall$ & $\LHI\sigstart, \ST\sigwall$\\
$\RLO\sigstart, \LHI\sigstart$ & $\ST\sigstart$
\end{tabular}\\\noalign{\vskip5mm}%
\begin{tabular}{r@{\texttt{@}}l}
\multicolumn{2}{c}{\textbf{Initial configuration}}\\\hline\noalign{\vskip2mm}%
\qquad$\SET{\ST\sigwall,\RLO\sigstart,\RHI\sigstart}$ & 0\\
$\SET{\ST\sigwall}$ & 1
\end{tabular}
\end{tabular}
\caption{\label{fig:middle}Geometrical algorithm for computing the middle}
\end{figure}

\paragraph{Signals.}
Each \emph{signal} is an instance of a \emph{meta-signal}.  The associated
meta-signal defines its \emph{velocity} and what happen when signals meet.
\Figure{fig:middle} presents a very simple space-time diagram.  Time is
increasing upwards and the meta-signals are indicated as labels on the signals.

Generally, we use over-line arrows to indicate the direction and speed of
propagation of a meta-signal.  For example, $\RLO\sigstart$ and $\LHI\sigstart$
denotes two different meta-signals; the first travels to the right at speed 1,
while the other travels to the left at speed $-3$. $\ST\sigwall$ and
$\ST\sigstart$ are both stationary meta-signals.

\paragraph{Collision rules.}
When a set of signals collide, they are replaced by a new set of signals
according to a matching collision rule.  A rule has the form:
\[
\sigma_1,\ldots,\sigma_n \rightarrow \sigma'_1,\ldots,\sigma'_p
\]
where all $\sigma_i$ are meta-signals of distinct speeds as well as $\sigma'_j$
(two signals cannot collide if they have the same speed and outcoming signals
must have different speeds).  A rule matches a set of colliding signals if its
left-hand side is equal to the set of their meta-signals.  By default, if there
is no exactly matching rule for a collision, the behavior is defined to
regenerate exactly the same meta-signals.  In such a case, the collision is
called \emph{blank}.  Collision rules can be deduced from space-time diagrams
as on \Fig{fig:middle}.  They are also listed on the right of this figure.

\paragraph{Signal machine.}
A signal machine is defined by a set of meta-signals, a set of collision rules,
and and initial configuration, i.e.\ a set of particles placed on the real
line.  The evolution of a signal machine can be represented geometrically as a
\emph{space-time diagram}: space is always represented horizontally, and time
vertically, growing upwards.  The geometrical algorithm displayed in
\Fig{fig:middle} computes the middle: the new $\ST\sigstart$ is located exactly
halfway between the initial two $\ST\sigwall$.

%
\section{Computing in the fractal cloud}
\label{sec:fractal-cloud}

\paragraph{Constructing the fractal:}
the fractal structure that interests us is based on the simple idea of
computing the middle illustated in Figure~\ref{fig:middle}.  We just
indefinitely repeat this geometrical construction: once space has been halved,
we recursively halve the two halves, and so on.
\begin{figure}[hbt]
\vspace*{-7em}%
\centering
\subfigure[Constructing the fractal cloud\label{fig:fractal:basic}]{%
  \includegraphics[width=0.4\linewidth]{tree.pdf}}
\quad
\subfigure[Distributing a computation\label{fig:fractal:distrib:bug}]{%
  \includegraphics[width=0.4\linewidth]{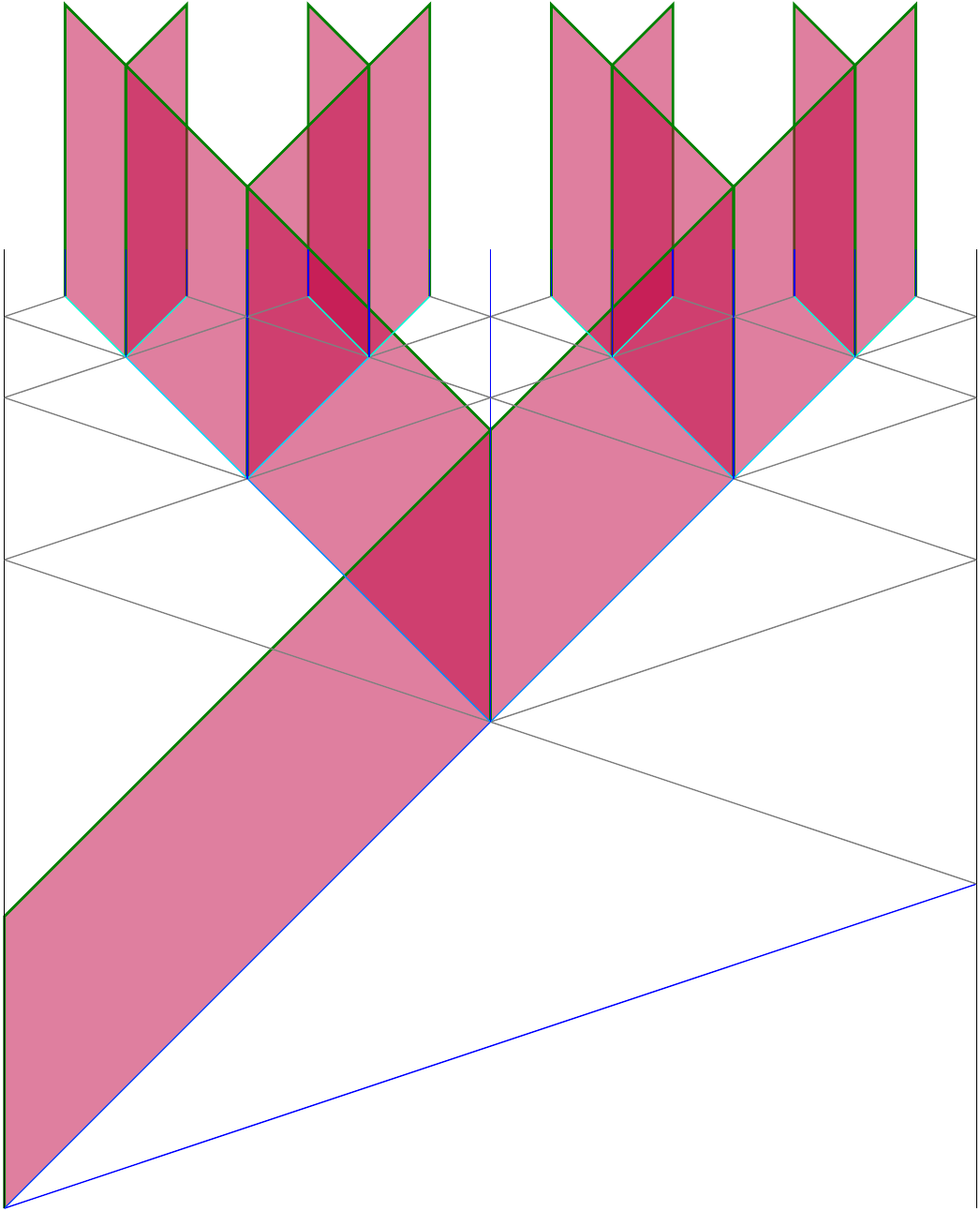}}
\caption{Computing in the fractal cloud}
\end{figure}
This is illustrated in Figure~\ref{fig:fractal:basic}, and can be generated by
the following rules:\footnote{For brevity, we will always omit the rules which can
  be obtained from the others by symmetry. We refer to \Appendix{sec:appendixA} for more details.}
\begin{align*}
\sigwall,\LHI\sigstart & \becomes \sigwall,\RHI\sigstart &
\RLO\sigstart, \LHI\sigstart & \becomes
\LHI\sigstart, \LLO\sigstart, \ST\sigstart, \RLO\sigstart, \RHI\sigstart
\end{align*}
using $\AT{\ST\sigwall,\RLO\sigstart,\RHI\sigstart}0$ $\AT{\ST\sigwall}1$ as
the initial configuration.  This produces a stack of levels: each level is half
the height of the previous one.  As a consequence, the full fractal has width 1
and height 1.

\paragraph{Distributing a computation:}
the point of the fractal is to recursively halve space.  At each point where
space is halved, we position a stationary signal (a vertical line in the
space-time diagram).  We can use this structure, so that, at each halving point
(stationary signal), we split the computation in two: send it to the left with
half the data, and also to the right with the other half of the data.

The intuition is that the computation is represented by a \emph{beam} of
signals, and that stationary signals split this beam in two, resulting in one
beam that goes through, and one beam that is reflected.

Unfortunately, a beam of constant width will not do: eventually it becomes too
large for the height of the level.  This can be clearly seen in
Figure~\ref{fig:fractal:distrib:bug}.

\paragraph{The lens device:}
the lens device narrows the beam by a factor of 2 at each level, thus
automatically adjusting it to fit the fractal.  It is implemented by the
following meta-rule: \textit{unless otherwise specified, any signal $\RLO\sigma$
  is accelerated by $\LHI\sigstart$ and decelerated and split by any stationary
  signal $\ST\sigs$.}
\begin{figure}[htb]
\centering
\subfigure[narrows by 2]{%
\begin{tikzpicture}[x=2em,y=2em]
\draw[dotted] (-9,0) -- (1,0);
\draw[->] (0,-1) -- (0,5);
\draw (-9,-1) -- node[above] {$\overrightarrow\sigma$} (-6,2);
\draw (-6,2) -- node[above] {$\Overrightarrow\sigma$} (0,4);
\draw[->] (0,4) -- node[above] {$\overleftarrow\sigma$} (-1,5);
\draw[->] (0,4) -- node[above] {$\overrightarrow\sigma$} (1,5);
\draw[->] (0,0) -- node[above] {$\LHI\sigstart$} (-6,2) -- (-9,3);
\draw[dotted] (-6,0) -- (-6,2);
\draw[dotted] (-6,2) -- (0,2);
\draw[<->,gray] (-8,-0.2) -- node[below] {$t$} (-6,-0.2);
\draw[<->,gray] (-6,-0.2) -- node[below] {$3t$} (0,-0.2);
\draw[<->,gray] (0.2,0) -- node[right] {$t$} (0.2,2);
\draw[<->,gray] (0.2,2) -- node[right] {$t$} (0.2,4);
\end{tikzpicture}}
\quad
\subfigure[effect on propagation]{%
\mbox{\includegraphics[width=0.4\linewidth]{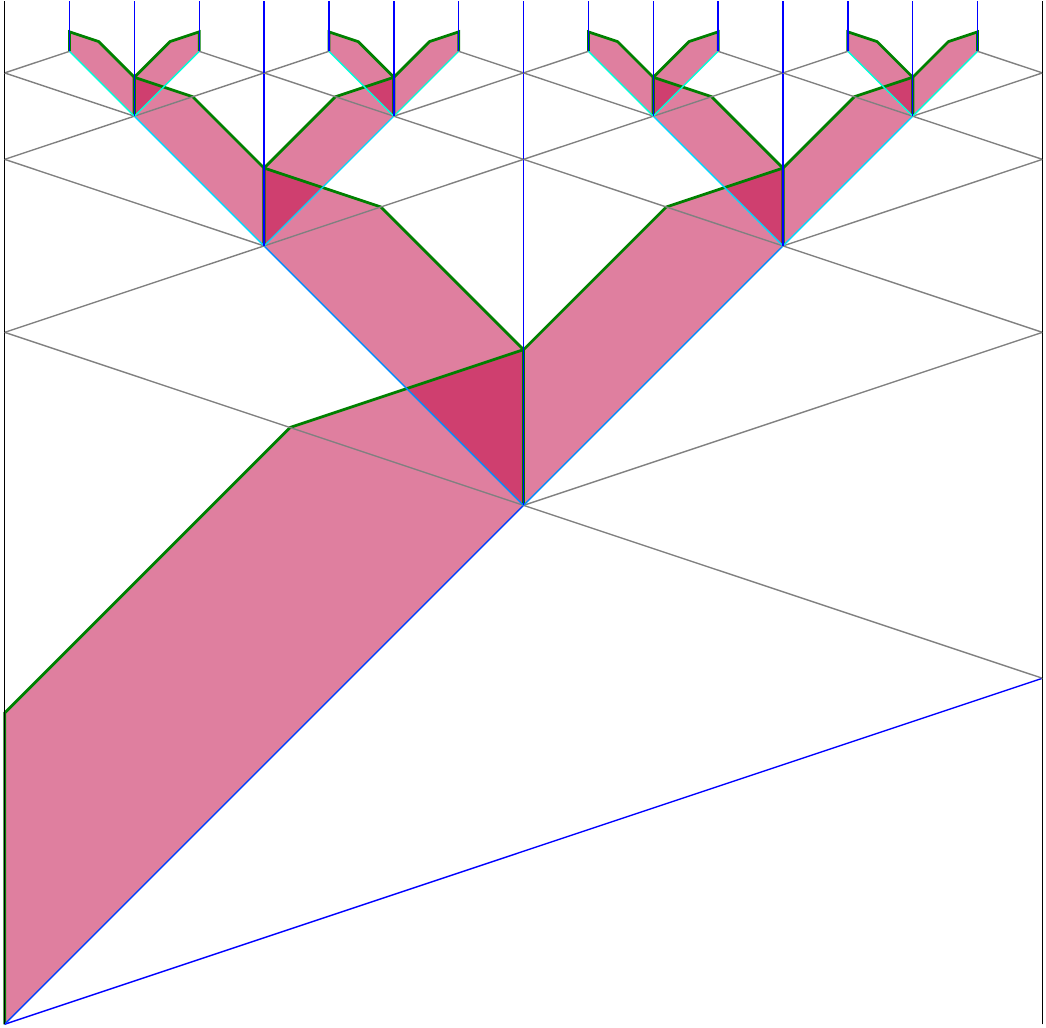}}}
\caption{\label{fig.lens}The lens device}
\end{figure}
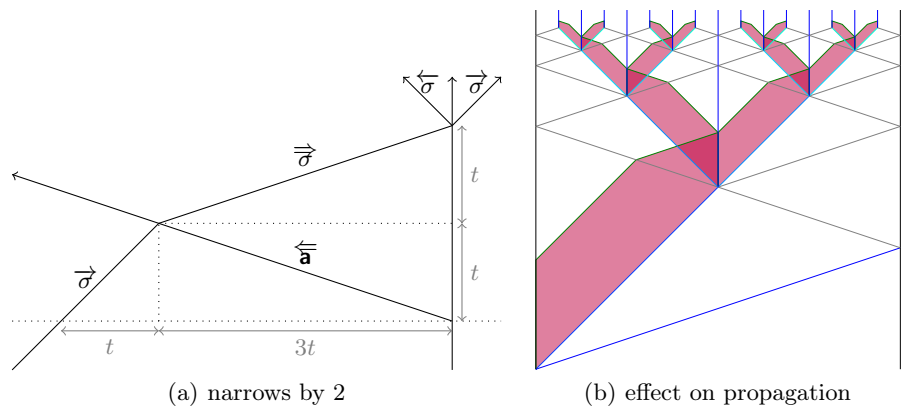

\paragraph{Generic computing over the fractal cloud:}
with the lens device in effect, generic computations can take place over the
fractal by propagating a beam from an initial configuration.  We write
$\block[\RLO{\sigma_n}\ldots\RLO{\sigma_1}]{spawn}$ for an initial
configuration with a sequence $\RLO{\sigma_n}\ldots\RLO{\sigma_1}$ of signals.
Geometrically, it can be seen easily that, in order for the beam to fit through
the first level, the sequence $\RLO{\sigma_n}\ldots\RLO{\sigma_1}$ must be
placed in the interval $(-\frac 1 4,0)$.

\paragraph{Stopping the fractal:}
For finite computations, we don't need the entire fractal.  The
$\block{until}[n]$ module can be inserted in the initial configuration to cut
the fractal after $n$ levels have been generated.\hfill
$\block{until}[n]=\RLO\sigstop{\RLO\sigstopaux}^{n-1}$
\begin{align*}
\RHI\sigstopaux, \LLO\sigstart & \becomes \LLO\sigstartoff,\RHI\sigstopaux &
\RHI\sigstopaux, \ST\sigstart & \becomes \ST\sigstartoff &
\RHI\sigstopaux, \ST\sigstartoff & \becomes \LLO\sigstopaux, \ST\sigstartoff, \RLO\sigstopaux\\
\RHI\sigstop, \LLO\sigstart & \becomes \LLO\sigstartoff, \RHI\sigstop &
\RHI\sigstop, \LLO\sigstartoff & \becomes \LLO\sigstart, \RHI\sigstop &
\RHI\sigstop, \ST\sigstartoff & \becomes \LLO\sigstop, \ST\sigstart, \RLO\sigstop\\
\RHI\sigstop, \ST\sigstart & \becomes \ST\sigstart, \RHI\sigstop &
\RHI\sigstop, \RLO\sigstart & \becomes \RLO\sigstartoff &
\RHI\sigstart, \LLO\sigstartoff & \becomes
\end{align*}
The subbeam ${\RLO\sigstopaux}^{n-1}$ are the inhibitors for $\RLO\sigstop$.  One
inhibitor is consumed at each level, after which $\RHI\sigstop$ takes effect
and turns $\LLO\sigstart$ into $\LLO\sigstartoff$, then crosses $\ST\sigstart$
and turns $\RLO\sigstart$ on the other side into $\RLO\sigstartoff$.  Finally
the annihilation rule $\RHI\sigstart, \LLO\sigstartoff \becomes \emptyset$
brings the fractal to a stop.  Thus, a computation
$\block[\RLO{\sigma_n}\ldots\RLO{\sigma_1}{\block{until}[n]}]{spawn}$ uses only
$n$ levels.  It can be seen geometrically that, for the collision of
$\RHI\sigstop$ with $\RLO\sigstart$ to occur before the latter meets with
$\LHI\sigstart$, $\RLO\sigstop$ must initially be placed in $(-\frac 1 6,0)$.

%
\section{A modular Q-SAT solver}
\label{sec:qsat}

Q-SAT is the satisfiability problem for quantified Boolean formulae (QBF).  A
QBF is a closed formula of the form: \[\phi~=~Q_1x_1Q_1x_2\ldots
Q_nx_n~~~\psi(x_1,x_2,\ldots,x_n)\] where $Q_i\in\{\exists,\forall\}$ and
$\psi$ is a quantifier-free formula of propositional logic.  A recursive
algorithm for solving Q-SAT is:
\begin{align*}
\qsat(\exists x\ \phi) & = \qsat(\phi[x\leftarrow\qfalse]) \vee
\qsat(\phi[x\leftarrow\qtrue])\\
\qsat(\forall x\ \phi) & = \qsat(\phi[x\leftarrow\qfalse]) \wedge
\qsat(\phi[x\leftarrow\qtrue])\\
\qsat(\beta) & = \qeval(\beta)
\end{align*}
where $\beta$ is a ground Boolean formula.  This is exactly the structure of
our construction: each quantified variable splits the computation in 2,
$\qsat(\phi[x\leftarrow\qfalse])$ is sent to the left and
$\qsat(\phi[x\leftarrow\qtrue])$ to the right, and subsequently the recursively
computed results that come back are combined (with $\vee$ for $\exists$ and
$\wedge$ for $\forall$) to yield the result for the quantified formula.  This
process can be viewed as an instance of \emph{Map/Reduce}, where the \emph{Map}
phase distributes the combinatorial exploration of all possible valuations
across space using a binary decision tree, and the \emph{Reduce} phase collects
the results and aggregates them using quantifier-appropriate Boolean
operations.
Our Q-SAT solver is modularly composed as follows:
\[
\hss\block[{\block{reduce:qsat}[Q_1x_1\ldots Q_nx_n]
    \block{map:sat}[\psi]\block{decide}[n]\block{until}[n+1]}]{spawn}\hss
\]
We describe the modules \texttt{decide}, \texttt{map:sat}, and
\texttt{reduce:qsat} below.

\subsection{Setting up the decision tree}
For a QBF with $n$ variables, we need 1 level per variable, and then at level
$n+1$ we have a ground propositional formula that needs to be evaluated.  Thus,
the first module we insert is $\block{until}[n+1]$ to create $n+1$ levels.  We
then insert $\block{decide}[n]$ because we want to use the first $n$ levels as
decision points for each variable.\hfill
$\block{decide}[n] = {\RLO\sigstartaux}^n$
\begin{align*}
\RHI\sigstartaux, \ST\sigstart & \becomes \ST\sigx &
\RHI\sigstartaux, \ST\sigx & \becomes \LLO\sigstartaux, \ST\sigx, \RLO\sigstartaux
\end{align*}

\subsection{Compiling the formula}
The intuition is that we want to compile the formula into a form of inverse
polish notation to obtain executable code using postfix operators.  At level
$n+1$ all variables have been decided, and have become $\RLO\sigt$ or
$\RLO\sigf$. The ground formula, regarded as an expression tree, can be
executed bottom up to compute its truth value: the resulting signal for a
subexpression is sent to interact with its parent operator.

The formula is represented by a beam of signals: each subformula is represented
by a (contiguous) subbeam.  A subformula that arrives at level $n+1$ starts
evaluating when it hits the stationary $\ST\sigstart$.  When its truth value
has been computed, it is reflected so that it may eventually collide with the
incoming signal of its parent connective.

\paragraph{Compilation.}
For binary connectives: one argument arrives first, it is evaluated, and its
truth value is reflected toward the incoming connective; but, in order to reach
it, it must cross the incoming beam for the other argument and not interact
with the connectives contained therein.  For this reason, with each
subexpression, we associate a beam ${\RLO\siggamma}^k$ of inhibitors that
prevents its resulting truth value to interact with the first $k$ connectives
that it crosses.  We write $\CC{\psi}$ for the compilation of $\psi$
into a contribution to the initial configuration, and $\NCON{\psi}$ for the
number of occurrences of connectives in $\psi$.
\begin{align*}
\CC{\psi} &= \CC{\psi}^0\\
\CC{\psi_1\AND\psi_2}^k &= \RLO\sigand\ {\RLO\siggamma}^k\ \CC{\psi_1}^0\ \CC{\psi_2}^{\NCON{\psi_1}}\\
\CC{\psi_1\OR\psi_2}^k &= \RLO\sigor\ {\RLO\siggamma}^k\ \CC{\psi_1}^0\ \CC{\psi_2}^{\NCON{\psi_1}}\\
\CC{\NOT\psi}^k &= \RLO\signot\ {\RLO\siggamma}^k\ \CC{\psi}\\
\CC{x_i}^k & = \block{var}[x_i]\ {\RLO\siggamma}^k
\end{align*}

\paragraph{Variables.}
We want variable $x_i$ to be decided at level $i$.  This can be achieved using
$i-1$ inhibitors.\hfill
$\block{var}[x_i] = \RLO\sigx{\RLO\sigxdelay}^{i-1}$
\begin{align*}
\RHI\sigxdelay, \ST\sigx & \becomes \ST\sigxoff &
\RHI\sigxdelay, \ST\sigxoff & \becomes \LLO\sigxdelay, \ST\sigxoff, \RLO\sigxdelay \\
\RHI\sigx, \ST\sigx & \becomes \LLO\sigf, \ST\sigx, \RLO\sigt &
\RHI\sigx, \ST\sigxoff & \becomes \LLO\sigx, \ST\sigx, \RLO\sigx
\end{align*}
For variable $x_i$, the idea is to protect $\RHI\sigx$ from being assigned into
$\LLO\sigf$ and $\RLO\sigt$ until it reaches the $i^{th}$ level. This is
achieved with a stack of $i-1$ signals $\RHI\sigxdelay$: at each level, the
first $\RHI\sigxdelay$ turns the stationary signal $\sigx$ into $\sigxoff$(the
non-assigning version of $\ST\sigx$), and dies.  The following $\RHI\sigxdelay$
are simply split, and so is $\RHI\sigx$ but it additionally turns $\ST\sigxoff$
back into $\ST\sigx$. After the first $i-1$ levels, all the $\RLO\sigxdelay$
have been consumed so that $\RHI\sigx$ finally collides with $\sigx$ and splits
into $\LLO\sigf$ going left and $\RLO\sigt$ going right.

\begin{figure}[tb]\centering
\subfigure[\label{fig:eval}Evaluation 
case~$x_1=x_2=x_3=\sigt$]{%
\includegraphics[width=0.4\linewidth]{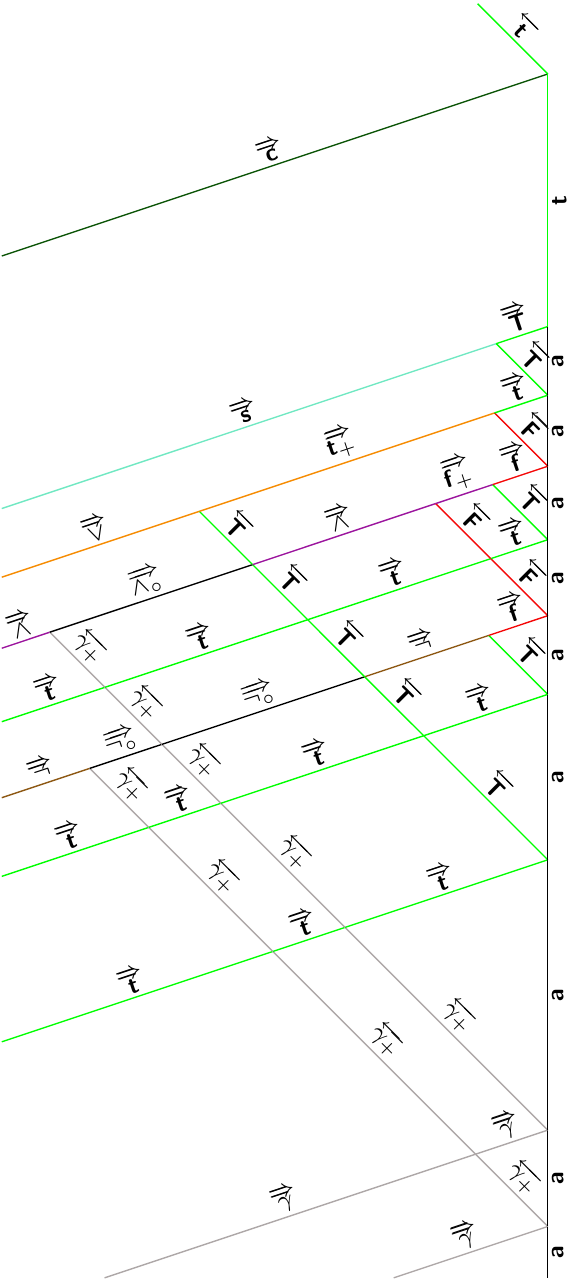}}
\hfill
\subfigure[\label{fig:aggreg}Aggregation]{%
\begin{minipage}[b]{0.58\linewidth}
\raggedleft
\mbox{$\block[\RLO\sigcollect\RLO\sigforall\RLO\sigforall\RLO\sigexists
\RLO\sigs\RLO\sigor\RLO\sigand\RLO\sigx\RLO\signot\RLO\sigx
\RLO\sigxdelay\RLO\sigx\RLO\sigxdelay\RLO\sigxdelay
\RLO\siggamma\RLO\siggamma
\RLO\sigstartaux\RLO\sigstartaux\RLO\sigstartaux
\RLO\sigstop\RLO\sigstopaux\RLO\sigstopaux\RLO\sigstopaux]{spawn}
$\hss}
\centerline{\textbf{Initial configuration}}
\vspace*{5mm}
\begin{align*}
\RHI\sigcollect, \ST\sigt & \becomes \LLO\sigt &
\RHI\sigcollect, \ST\sigf & \becomes \LLO\sigf\\[2mm]
\RLO\sigt, \ST\sigexistsL, \LLO\sigt & \becomes \LLO\sigt &
\RLO\sigt, \ST\sigforallL, \LLO\sigt & \becomes \LLO\sigt \\
\RLO\sigt, \ST\sigexistsL, \LLO\sigf & \becomes \LLO\sigt &
\RLO\sigt, \ST\sigforallL, \LLO\sigf & \becomes \LLO\sigf \\
\RLO\sigf, \ST\sigexistsL, \LLO\sigt & \becomes \LLO\sigt &
\RLO\sigf, \ST\sigforallL, \LLO\sigt & \becomes \LLO\sigf \\
\RLO\sigf, \ST\sigexistsL, \LLO\sigf & \becomes \LLO\sigf &
\RLO\sigf, \ST\sigforallL, \LLO\sigf & \becomes \LLO\sigf
\end{align*}
\includegraphics[width=\linewidth]{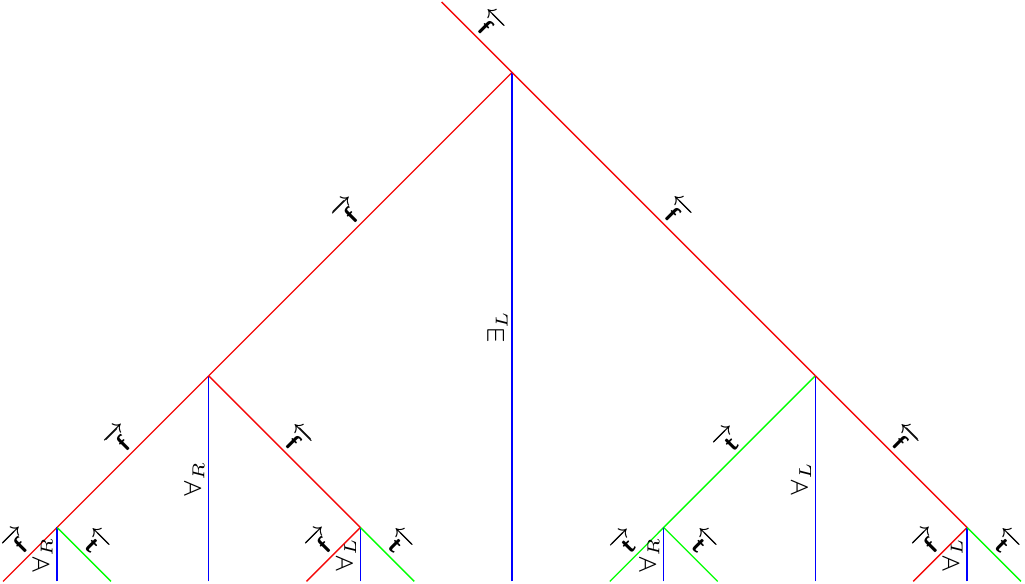}
\end{minipage}}
\caption{Example $\exists x_1\forall x_2\forall
  x_3\ (x_1\AND \neg x_2)\OR x_3$}
\end{figure}

\paragraph{Evaluation.}
When hitting $\ST\sigstart$ at level $n+1$, $\RHI\sigt$ is reflected as
$\LLO\sigT$, and $\RHI\sigf$ as $\LLO\sigF$: these are their \emph{activated}
versions which can interact with incoming connectives to compute the truth
value of the formula according to the rules below (for $\AND$; other
connectives are similar, \emph{cf} \Appendix{sec:appendixA}).  See Fig.~\ref{fig:eval} for an example.
\begin{align*}
\RHI\sigt, \ST\sigstart & \becomes \LLO\sigT, \ST\sigstart &
\RHI\sigf, \ST\sigstart & \becomes \LLO\sigF, \ST\sigstart &
\RHI\siggamma, \ST\sigstart & \becomes \LLO\siggammaP, \ST\sigstart\\
\RHI\sigand, \LLO\sigT & \becomes \RHI\sigandP &
\RHI\sigfP, \LLO\sigT & \becomes \RHI\sigf &
\RHI\sigandP, \LLO\sigT & \becomes \RHI\sigt\\
\RHI\sigand, \LLO\sigF & \becomes \RHI\sigfP &
\RHI\sigfP, \LLO\sigF & \becomes \RHI\sigf &
\RHI\sigandP, \LLO\sigF & \becomes \RHI\sigf\\
\RHI\sigand, \LLO\siggammaP & \becomes \RHI\sigandZ &
\RHI\sigandZ, \LLO\sigT & \becomes \LLO\sigT, \RHI\sigand &
\RHI\sigandZ, \LLO\sigF & \becomes \LLO\sigF, \RHI\sigand
\end{align*}

\paragraph{Storing the result.}
In order to make the result easily exploitable by the \emph{Reduce} phase, we
now store it as the stationary signal at level $n+1$; it replaces
$\ST\sigstart$.
\hspace*{\fill}$\block{store} = \RLO\sigs$
\begin{align*}
\RHI\sigs, \LLO\sigT & \becomes \RHI\sigT &
\RHI\sigs, \LLO\sigF & \becomes \RHI\sigF &
\RHI\sigT, \ST\sigstart & \becomes \ST\sigt &
\RHI\sigF, \ST\sigstart & \becomes \ST\sigf
\end{align*}
The complete \emph{Map} phase in implemented by:\hfill
$\block{map:sat}[\psi] = \block{store}\CC{\psi}$

\subsection{Aggregating the results}
As explained earlier, the results for an existentially (resp.\ universally)
quantified variable must be combined using $\OR$ (resp.\ $\AND$).

\paragraph{Setting up the quantifiers.}
We turn the decision points of the first $n$ levels into quantifier signals.
Moreover, at each level, we must also take note of the direction in which the
aggregated result must be sent.  Thus $\ST\sigexistsL$ represents an
existential quantifier that must send its result to the left.\EOL
\hspace*{\fill}$\block{reduce:qsat:init}[Q_1x_1 \cdots Q_nx_n] = \RLO{Q_n}\ldots\RLO{Q_1}$
\begin{align*}
\ST\sigx, \LHI\sigexists & \becomes \ST\sigexistsR & 
\RHI\sigexists, \ST\sigx & \becomes \ST\sigexistsL &
\ST\sigx, \LHI\sigforall & \becomes \ST\sigforallR &
\RHI\sigforall, \ST\sigx & \becomes \ST\sigforallL
\end{align*}

\paragraph{Aggregating the results.}
Actual aggregation is initiated by $\RLO\sigcollect$ and then executes
according to the rules given in Fig~\ref{fig:aggreg}.
\hspace*{\fill}$\block{reduce:qsat:exec}=\RLO\sigcollect$

\paragraph{The complete \emph{Reduce} phase} is implemented by
\hspace*{\fill}$\block{reduce:qsat}[Q_1x_1 \cdots Q_nx_n] =
\block{reduce:qsat:exec}\block{reduce:qsat:init}[Q_1x_1 \cdots Q_nx_n]$

%
\section{Machines for SAT variants}
\label{sec:sat-variants}

Similar machines for variants of SAT can be obtained easily, typically by using
different modules for the \emph{Reduce} phase.

\paragraph{ENUM-SAT.}
returning all the satisfying assignments for a propositional formula $\psi$ can
be achieved easily by storing them as stationary beams.\EOL
\hspace*{\fill}$\block{reduce:allsat}[n] =
\RLO\sigv\block{var}[x_1]\ldots\block{var}[x_n]\RLO\sigv$
\begin{align*}
\RHI\sigv, \ST\sigt & \becomes \LLO\sigvO, \ST\sigv &
\RLO\sigvO, \LHI\sigt & \becomes \ST\sigt, \RLO\sigvO &
\RHI\sigv, \ST\sigf & \becomes \LLO\sigvZ &
\RLO\sigvZ, \LHI\sigt & \becomes \RLO\sigvZ\\
\RLO\sigvO, \LHI\sigv & \becomes \ST\sigv &
\RLO\sigvO, \LHI\sigf & \becomes \ST\sigf, \RLO\sigvO &
\RLO\sigvZ, \LHI\sigv & \becomes &
\RLO\sigvZ, \LHI\sigf & \becomes \RLO\sigvZ
\end{align*}

\paragraph{\#SAT.}
counting the number of satisfying assignments for $\psi$ can be achieved using
signals for a binary adder.  For lack of space, we cannot exhibit the rules
here, but they can be found in \Appendix{sec:appendixB}.

\paragraph{MAX-SAT.}
finding the maximum number of \emph{clauses} that can be satisfied by an
assignment.  Here we must count the number of satisfied clauses rather than the
number of satisfying assignments, and then stack a module for computing the max
of two binary numbers.

%
\section{Complexities}
\label{sec:complexities}

As mentioned in \Sec{sec:intro}, we implement algorithms for
satisfiability problems on signal machines in order to
investigate the computational power of our abstract geometrical
model of computation and to compare it to others.  As we shall see, for such
comparisons to be meaningful, the way complexity is measured must be adapted to
the nature on the computing machine.

\begin{wrapfigure}{l}{.45\textwidth}\vspace*{-7mm}%
  \centering
  \includegraphics[width=.45\textwidth]{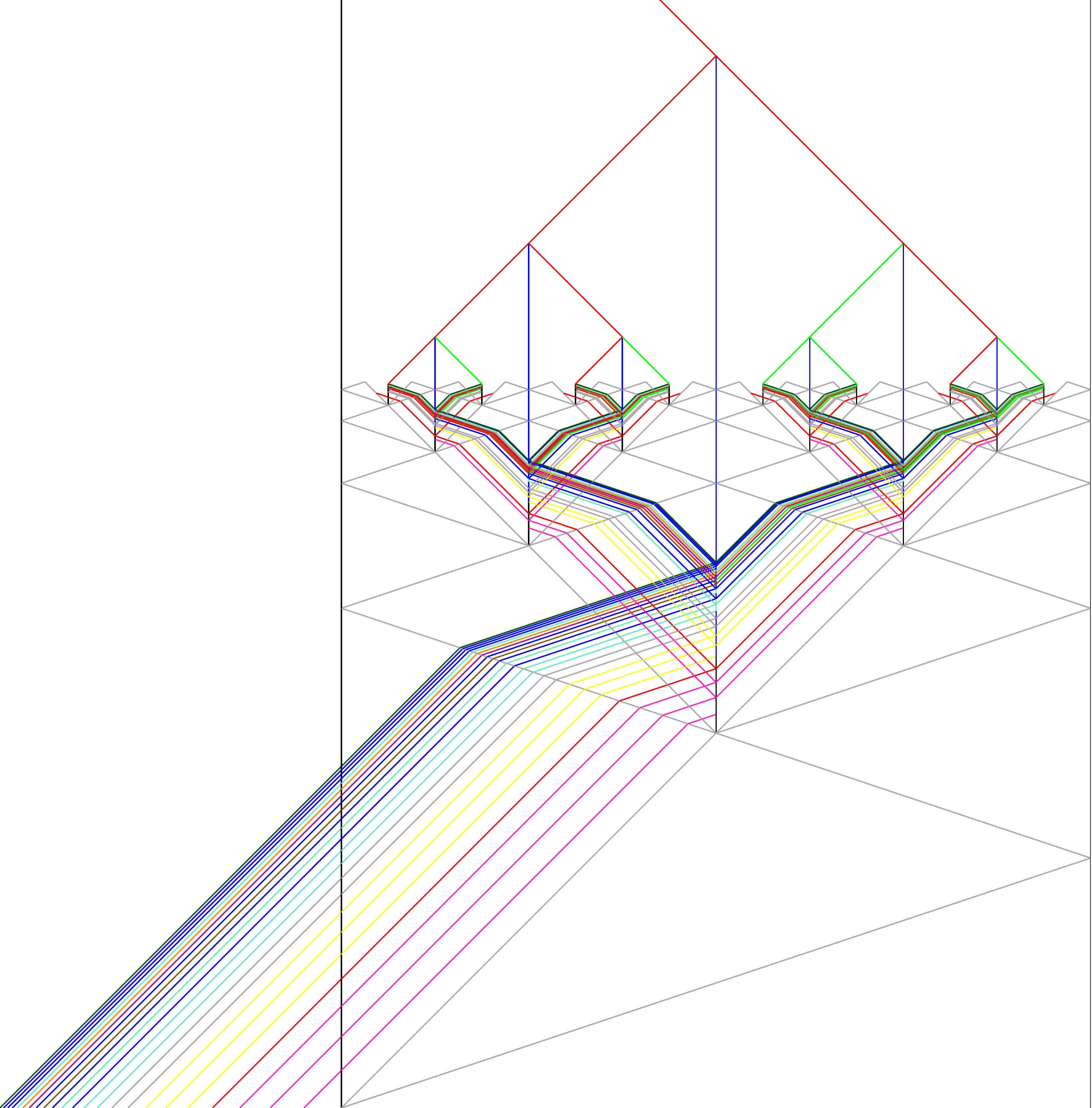}
  \caption{The whole diagram.}
  \label{fig:whole_diagram}
  \vspace*{-6.5mm}%
\end{wrapfigure}

Since signal machines can be regarded as the extension of cellular automata
from discrete to continous time and space, it might seem natural to measure the
time (resp.\ space) complexity of a computation using the height (resp.\ width)
of its space-time diagram.  But, in our applications to SAT variants, these are
bounded and independant of the formula: the \emph{Map} phase is bounded by the
fractal, and, by symmetry, so is the \emph{Reduce} phase.  Indeed, in general,
by an appropriate scaling of the initial configuration, a finite computation
could be made as small as desired.  Thus, height and width are no longer
pertinent measures of complexity.

Instead, we should regard our construction as a massively parallel
computational device transforming inputs into outputs.  The input is the
initial configuration at the bottom of the diagram, and the output is the truth
value signal coming out at the top, as seen in~\Fig{fig:whole_diagram} for
formula $\exists x_1\forall x_2\forall x_3\ (x_1\AND\neg x_2)\OR x_3$.  The
transformation is performed in parallel by many threads: a thread here is an
ascending path through the diagram from an input to the output, and the
operations executed by the thread are the collisions occurring on this path.

Formally, we view a space-time diagram as a directed acyclic graph of
collisions (vertices) and signals (arcs) oriented according to causality.  Time
complexity is then defined as the maximal size of a chain of collisions \ie the
length of the longest path, and space complexity as the maximal size of an
anti-chain \ie the size of the maximal set of signals pairwise un-related.
This model-specific measure of time complexity is called \emph{collisions
  depth}.

For the present construction, if $s$ is the size of the formula and $n$ the
number of variables, space complexity is exponential: during evaluation, $2^n$
independent computations are executed in parallel, each one involving
approximately $s$ signals, so that the total space complexity is in $O(s.2^n)$.

Regarding the time complexity: for each subformula, the compilation process
introduces a number of signals at most linear in $s$.  Thus the number of
signals in the initial configuration is $O(s^2)$.  The primary contribution to
the number of collisions along an ascending path comes, at each of the $n$
levels, from the reflected beam crossing the incoming beam.  Thus a thread
involves $O(n.s^2)$ collisions, making the collision depth cubic in the size of
the formula instead of quadratic for our previous family of machines, giving us an
idea of the price for genericity.


%
\section{Conclusion}
\label{sec:conclusion}

We showed in this paper that abstract geometrical computation can solve Q-SAT
in bounded space and time by means of a single generic signal machine. This is
achieved through massive parallelism enabled by a fractal construction that we
call the \emph{fractal cloud}.  We adapted the Map/Reduce paradigm to this
fractal cloud, and described a modular programming approach making it easy to
assemble generic machines for SAT variants such as \#SAT or MAX-SAT.

As we explained in \Sec{sec:complexities}, time and space are no longer
appropriate measures of complexity for geometrical computations.  This leads us
to propose new definitions of complexity, specific to signal machines, and
taking in account the parallelism of the model: time and space complexities are
now defined respectively by the maximal sizes of a chain and an anti-chain,
when the diagram is regarded as a directed acyclic graph. Time complexity thus
defined is called \emph{collision depth}.  According to these new definitions,
our construction has exponential space complexity and cubic time complexity.

Although the model is purely theoretical and has no ambition to be physically
realizable, it is a significant and distinguishing aspect of signal machines
that they solve satifiability problems while adhering to major principles of
modern physics --- finite density and speed of information, causality --- that
are typically not considered by other unconventional models of computation.
They do not, however, respect the quantization hypothesis, nor the uncertainty
principle.

We are know furthering our research along two axes.  First, the design and
applications of other fractal structures for modular programming with fractal
parallelism.  Second, the investigation of computational complexity classes,
both classical and model-specific for abstract geometrical computation.



%


\small
\bibliographystyle{plainnat}


\bibliography{Duchier-Durand-Lose-Senot_2011}

\appendix

\newpage

\section{Details of the modules}
\label{sec:appendixA}

In the following, we generally define the collision rules only for one side,
as the corresponding rules for the other side can be deduced by symetry.
We give here the rules for all modules: first the modules for building and stopping
the fractal tree, then the modules for setting and using the tree for satisfiability
problems and finally the modules specific to each variants of SAT.

All the diagrams of the paper were generated by Durand-Lose's software, implemented in Java, 
and corresponds to a run of our Q-SAT solver for the formula:
\[
\phi = \exists x_1\forall x_2\forall x_3\ (x_1\AND \neg x_2)\OR x_3 \enspace.
\]

\subsection{The fractal cloud}

\begin{figure}[hbt]
  \centering
  \includegraphics[width=0.4\textwidth]{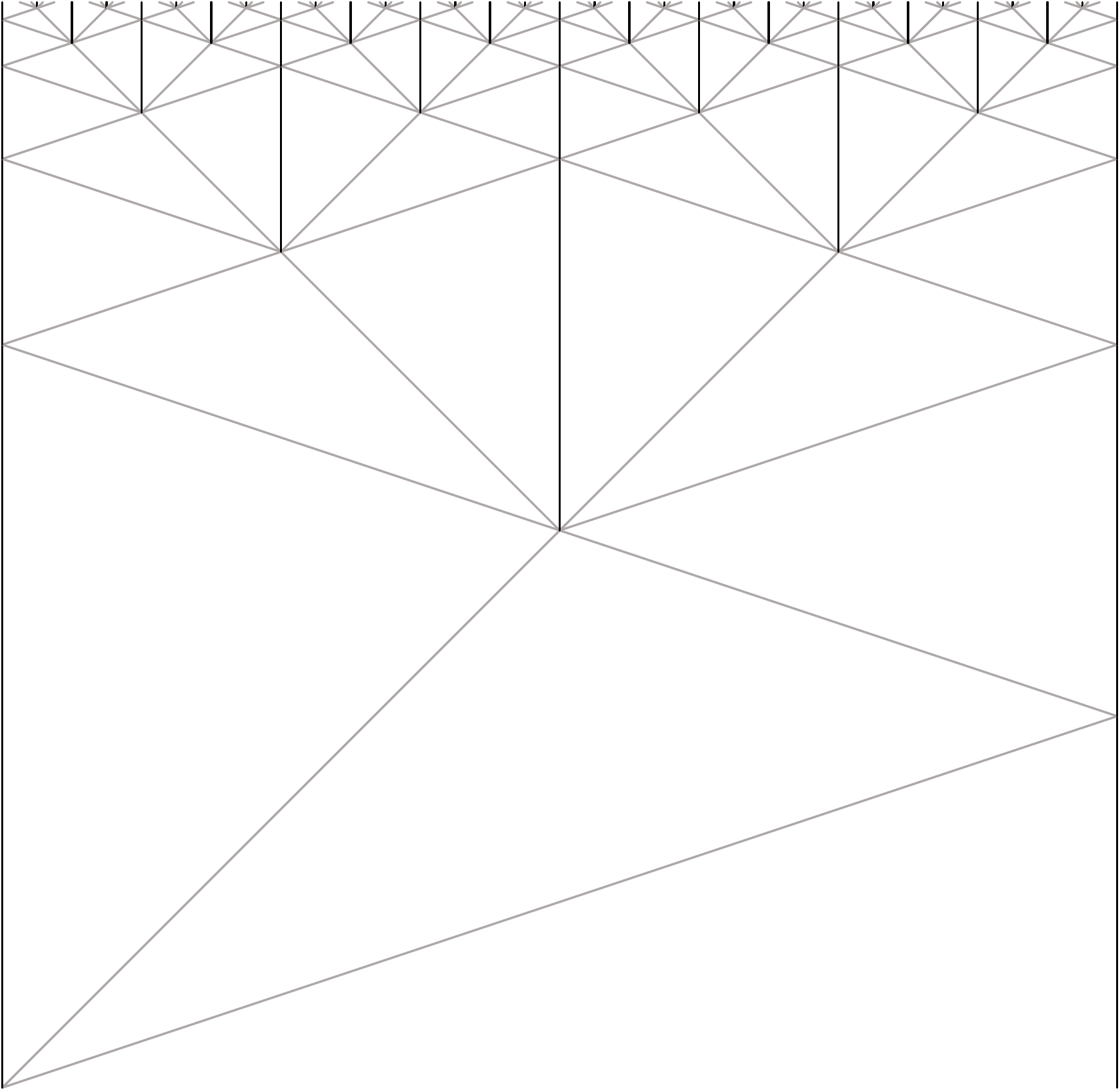}  
  \caption{The fractal tree.}
  \label{fig:app:tree}
\end{figure}

\paragraph{Constructing the fractal:}
$\texttt{[start]} = \RLO\sigstart,\RHI\sigstart$\\
The following rules on the left correspond to the bounce of $\RHI\sigstart$ and
$\LHI\sigstart$ on the wall $\ST\sigwall$. Rules on the right are the step of induction 
for starting the next level: the initial signals $\RLO\sigstart$ and $\RHI\sigstart$ are 
duplicated on the right and the left, and the stationary signal $\ST\sigstart$ is created
exactly at the middle of the previous stage. The result is given by \Fig{fig:app:tree}.
\begin{align*}
\sigwall,\LHI\sigstart & \becomes \sigwall,\RHI\sigstart &
\RLO\sigstart, \LHI\sigstart & \becomes \LHI\sigstart, \LLO\sigstart, \ST\sigstart, \RLO\sigstart, \RHI\sigstart\\
\RHI\sigstart, \sigwall & \becomes \LHI\sigstart, \sigwall &
\RHI\sigstart, \LLO\sigstart & \becomes \LHI\sigstart, \LLO\sigstart, \ST\sigstart, \RLO\sigstart, \RHI\sigstart\\
\end{align*}

\paragraph{Stopping the fractal:}
$\block{until}[n+1] = \RLO\sigstop{\RLO\sigstopaux}^{n}$

The role of this module is to stop the fractal after $(n+1)$ levels --- $n$
levels for assigning the $n$ variables and $1$ level for the evaluation of the
ground formula.  For this, we use a stack of $n$ signals $\RLO\sigstopaux$ and
one signal $\RLO\sigstop$. The $\RLO\sigstopaux$ signals are used both as a
counter, one signal being killed at each level, and as inhibitors to the effect
of $\RLO\sigstop$. After $n$ levels, only $\RLO\sigstop$ remains and can
stop the construction of the fractal at level $n+1$.  Here are the corresponding rules:
\begin{align*}
\RHI\sigstopaux, \LLO\sigstart & \becomes \LLO\sigstartoff,\RHI\sigstopaux &
\RHI\sigstopaux, \ST\sigstart & \becomes \ST\sigstartoff &
\RHI\sigstopaux, \ST\sigstartoff & \becomes \LLO\sigstopaux, \ST\sigstartoff, \RLO\sigstopaux\\
\RHI\sigstop, \LLO\sigstart & \becomes \LLO\sigstartoff, \RHI\sigstop &
\RHI\sigstop, \LLO\sigstartoff & \becomes \LLO\sigstart, \RHI\sigstop &
\RHI\sigstop, \ST\sigstartoff & \becomes \LLO\sigstop, \ST\sigstart, \RLO\sigstop\\
\RHI\sigstop, \ST\sigstart & \becomes \ST\sigstart, \RHI\sigstop &
\RHI\sigstop, \RLO\sigstart & \becomes \RLO\sigstartoff &
\RHI\sigstart, \LLO\sigstartoff & \becomes
\end{align*}

\paragraph{The lens effect:}

\begin{figure}[hbt]
  \centering
  \includegraphics[width=1\textwidth]{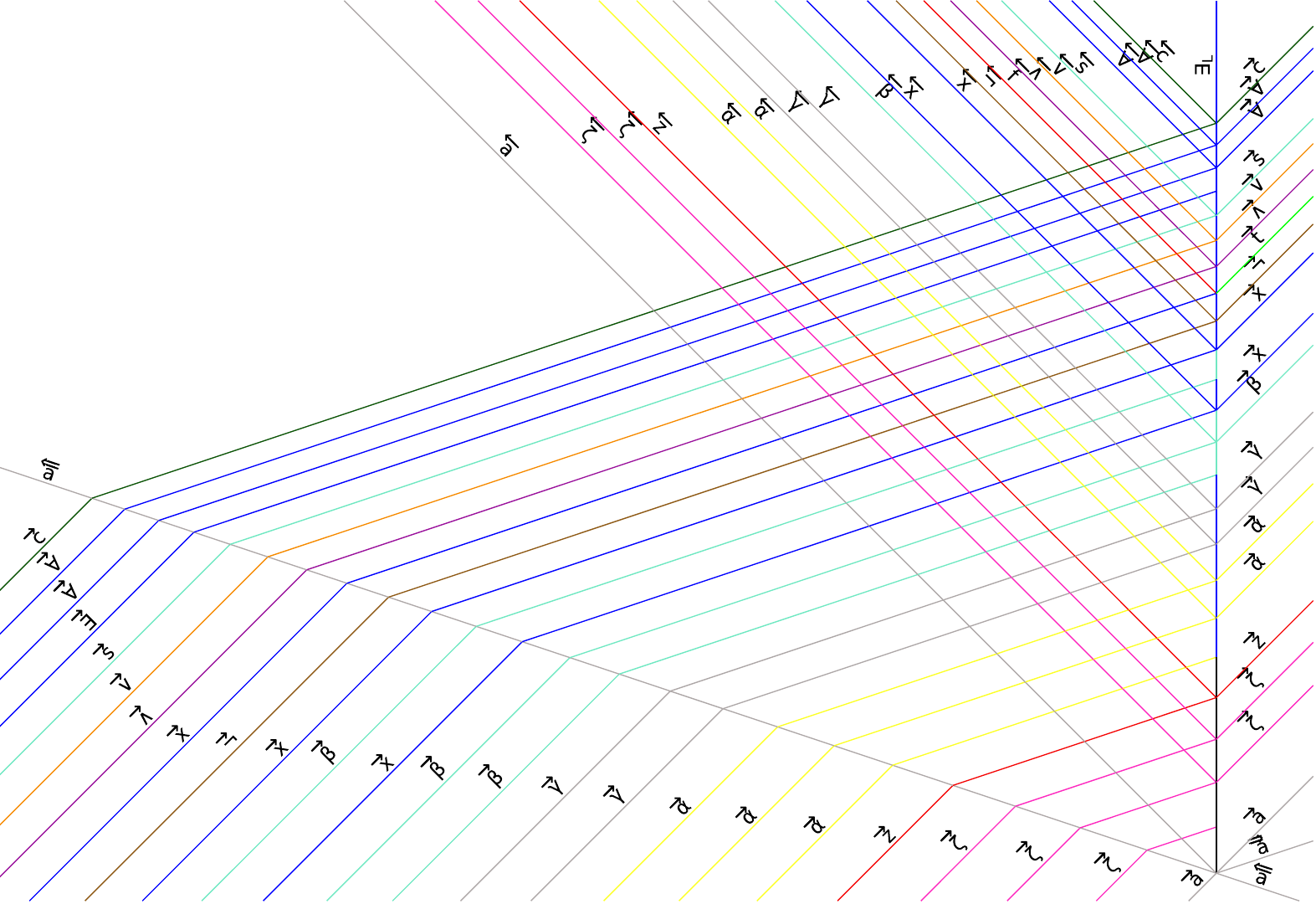}  
  \caption{Split and lens effect at the first level.}
  \label{fig:app:split}
\end{figure}

The general idea is that any signal $\RLO\sigi$ is accelerated by
$\LHI\sigstart$ and decelerated and split by any stationary signal
$\ST\sigs$. There are some special cases to handle for the deceleration and the
split, some particular signals being stopped at this moment. But in all cases,
signals are always accelerated by $\LHI\sigstart$. Figure \ref{fig:app:split}
zooms on a split and illustrates the lens effect on the beam as well as the
assignment of the top-most $\RLO\sigx$.

\emph{General case}: for any stationary signal $\ST\sigs$ (more exactly, $\ST\sigs$ is either $\ST\sigstart$,
$\ST\sigstartoff$, $\ST\sigx$, $\ST\sigxoff$, $\ST\sigexistsL$, $\ST\sigexistsR$,
$\ST\sigforallL$ or $\ST\sigforallR$) and for any signal $\RLO\sigi$ distinct
of $\RLO\sigstopaux$ and $\RLO\sigstop$, we have: 
\begin{align*}
\RLO\sigi, \LHI\sigstart & \becomes \LHI\sigstart, \RHI\sigi &
\RHI\sigi, \ST\sigs & \becomes \LLO\sigi, \ST\sigs, \RLO\sigi
\end{align*}

\emph{Case of $\RLO\sigstopaux$ and $\RLO\sigstop$}: the rules for applying the
lens effect to $\RLO\sigstopaux$ and $\RLO\sigstop$ are given previously in
§.Stopping the fractal.  The stationary signal involved here is $\ST\sigstart$,
which kills the first $\RHI\sigstopaux$, becomes $\ST\sigstartoff$ and splits
and decelerates the next signals $\RHI\sigstopaux$ and $\RHI\sigstop$, and then
becomes $\ST\sigstart$ again.

\emph{Case of $\RLO\sigstartaux$}: the first $\RLO\sigstartaux$ is stopped by the stationary
signal $\ST\sigstart$, which becomes $\ST\sigx$ and decelerates and splits the next coming
$\RLO\sigstartaux$. 
\begin{align*}
\RHI\sigstartaux, \ST\sigstart & \becomes \ST\sigx &
\RHI\sigstartaux, \ST\sigx & \becomes \LLO\sigstartaux, \ST\sigx, \RLO\sigstartaux
\end{align*}

\emph{Case of quantifiers signals}: the first quantifier signal, $\RHI\sigforall$ or $\RHI\sigexists$, 
colliding with a stationary $\ST\sigx$ is stopped and turns $\ST\sigx$ into 
$\ST\sigforallD$ or $\ST\sigexistsD$ ($D \in \{R, L\}$). This is achieved by the rules:
\begin{align*}
\ST\sigx, \LHI\sigexists & \becomes \ST\sigexistsR & 
\RHI\sigexists, \ST\sigx & \becomes \ST\sigexistsL &
\ST\sigx, \LHI\sigforall & \becomes \ST\sigforallR &
\RHI\sigforall, \ST\sigx & \becomes \ST\sigforallL
\end{align*}

 The next quantifier signals are decelerated and split by the 
new stationary signal $\ST\sigforallD$ or $\ST\sigexistsD$.
In the following rules, we have $D \in \{R, L\}$:
\begin{align*}
\RHI\sigexists, \ST\sigexistsD & \becomes \LLO\sigexists, \ST\sigexistsD, \RLO\sigexists & 
\RHI\sigexists, \ST\sigforallD & \becomes \LLO\sigexists, \ST\sigforallD, \RLO\sigexists\\
\RHI\sigforall, \ST\sigexistsD & \becomes \LLO\sigforall, \ST\sigexistsD, \RLO\sigforall & 
\RHI\sigforall, \ST\sigforallD & \becomes \LLO\sigforall, \ST\sigforallD, \RLO\sigforall\\
\end{align*}


\subsection{The tree for satisfiabilty problems}

\paragraph{Activation of the decision tree:}
$\block{init}[n] = {\RLO\sigstartaux}^n$
\begin{align*}
\RHI\sigstartaux, \ST\sigstart & \becomes \ST\sigx &
\RHI\sigstartaux, \ST\sigx & \becomes \LLO\sigstartaux, \ST\sigx, \RLO\sigstartaux
\end{align*}

\paragraph{Representation of a variable:}
$\block{var}[x_i] = \RLO\sigx{\RLO\sigxdelay}^{i-1}$

As we explained in the paper, the variable $x_i$ is represented by a stack of
$i$ signals: one signal $\RLO\sigx$ and $i-1$ signals $\RLO\sigxdelay$. The role 
of the signals $\RLO\sigxdelay$ is to protect $\RLO\sigx$ from being assigned 
before the $i^{th}$ level. The first signal $\RLO\sigxdelay$ of the stack is 
stopped at the next split, so that $i-1$ signals have disappeared just before
the $i^{th}$ stage. This is illustrated in \Fig{fig:app:split} and modelised
 by the following rules:

\begin{align*}
\RHI\sigxdelay, \ST\sigx & \becomes \ST\sigxoff &
\RHI\sigxdelay, \ST\sigxoff & \becomes \LLO\sigxdelay, \ST\sigxoff, \RLO\sigxdelay \\
\RHI\sigx, \ST\sigx & \becomes \LLO\sigf, \ST\sigx, \RLO\sigt &
\RHI\sigx, \ST\sigxoff & \becomes \LLO\sigx, \ST\sigx, \RLO\sigx
\end{align*}

\paragraph{Compilation of the formula:}

We propose here a recursive algorithm which takes as input an
unquantified formula --- a SAT-formula --- and outputs the part of
the initial configuration corresponding to the formula.
In the following schemes of compilation, $\NCON{\phi}$ designates
the number of occurences of Boolean connectives in formula
$\phi$.

\begin{align*}
\CC{\phi} &= \CC{\phi}^0\\
\CC{\phi_1\AND\phi_2}^k &= \RLO\sigand\ {\RLO\siggamma}^k\ \CC{\phi_1}^0\ \CC{\phi_2}^{\NCON{\phi_1}}\\
\CC{\phi_1\OR\phi_2}^k &= \RLO\sigor\ {\RLO\siggamma}^k\ \CC{\phi_1}^0\ \CC{\phi_2}^{\NCON{\phi_1}}\\
\CC{\NOT\phi}^k &= \RLO\signot\ {\RLO\siggamma}^k\ \CC{\phi}\\
\CC{x_i}^k & = \RLO\sigx\ {\RLO\sigbeta}^{i-1}\ {\RLO\siggamma}^k
\end{align*}

\paragraph{Evaluation:}

\begin{figure}
  \centering
  \includegraphics[width=0.7\textwidth]{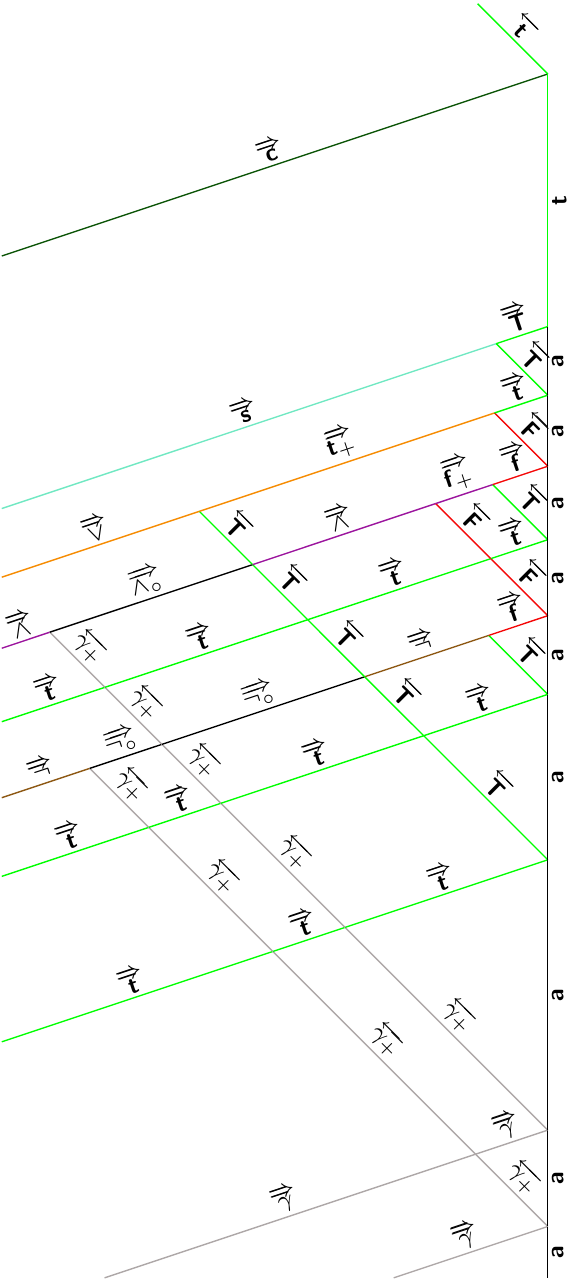}  
  \caption{Evaluation for $x_1=x_2=x_3=\ST\sigt$ in $\exists x_1\forall x_2\forall x_3\ (x_1\AND \neg x_2)\OR x_3$.}
  \label{fig:app:eval}
\end{figure}

The rules for the evaluation follow the classical Boolean operations. We explained earlier
that some inhibiting signals --- the $\RLO\siggamma$ --- are needed to allow
the result of the first evaluated argument of a binary connective to traverse
the beam of the other, as yet unevaluated, argument without reacting with the connectives contained
therein, and only interact with its actual syntactical parent connective.
Figure \ref{fig:app:eval} displays the evaluation of the 
formula $\phi=\exists x_1\forall x_2\forall x_3\ (x_1\AND \neg x_2)\OR x_3$ for the case
$x_1=x_2=x_3=\ST\sigt$.

\begin{align*}
\RHI\sigt, \ST\sigstart & \becomes \LLO\sigT, \ST\sigstart &
\RHI\sigf, \ST\sigstart & \becomes \LLO\sigF, \ST\sigstart &
\RHI\siggamma, \ST\sigstart & \becomes \LLO\siggammaP, \ST\sigstart\\[4mm]
\RHI\sigand, \LLO\sigT & \becomes \RHI\sigandP &
\RHI\sigfP, \LLO\sigT & \becomes \RHI\sigf &
\RHI\sigandP, \LLO\sigT & \becomes \RHI\sigt\\
\RHI\sigand, \LLO\sigF & \becomes \RHI\sigfP &
\RHI\sigfP, \LLO\sigF & \becomes \RHI\sigf &
\RHI\sigandP, \LLO\sigF & \becomes \RHI\sigf\\[4mm]
\RHI\sigor, \LLO\sigT & \becomes \RHI\sigtP &
\RHI\sigtP, \LLO\sigT & \becomes \RHI\sigt &
\RHI\sigorP, \LLO\sigT & \becomes \RHI\sigt\\
\RHI\sigor, \LLO\sigF & \becomes \RHI\sigorP &
\RHI\sigtP, \LLO\sigF & \becomes \RHI\sigt &
\RHI\sigorP, \LLO\sigF & \becomes \RHI\sigf\\[4mm]
\RHI\signot, \LLO\sigT & \becomes \RHI\sigf\\
\RHI\signot, \LLO\sigF & \becomes \RHI\sigt\\[4mm]
\RHI\sigand, \LLO\siggammaP & \becomes \RHI\sigandZ &
\RHI\sigor, \LLO\siggammaP & \becomes \RHI\sigorZ &
\RHI\signot, \LLO\siggammaP & \becomes \RHI\signotZ\\[4mm]
\RHI\sigandZ, \LLO\sigT & \becomes \LLO\sigT, \RHI\sigand &
\RHI\sigorZ, \LLO\sigT & \becomes \LLO\sigT, \RHI\sigor &
\RHI\signotZ, \LLO\sigT & \becomes \LLO\sigT, \RHI\signot\\
\RHI\sigandZ, \LLO\sigF & \becomes \LLO\sigF, \RHI\sigand &
\RHI\sigorZ, \LLO\sigF & \becomes \LLO\sigF, \RHI\sigor &
\RHI\signotZ, \LLO\sigF & \becomes \LLO\sigF, \RHI\signot
\end{align*}

\paragraph{Storing the results:} $\block{store} = \RLO\sigs$
\begin{align*}
\RHI\sigs, \LLO\sigT & \becomes \RHI\sigT &
\RHI\sigT, \ST\sigstart & \becomes \ST\sigt\\
\RHI\sigs, \LLO\sigF & \becomes \RHI\sigF &
\RHI\sigF, \ST\sigstart & \becomes \ST\sigf
\end{align*}

%

\section{Modularity and satifiability variants}
\label{sec:appendixB}

\subsection{Q-SAT}

\begin{figure}[hbt]
  \centering
  \includegraphics[width=0.9\textwidth]{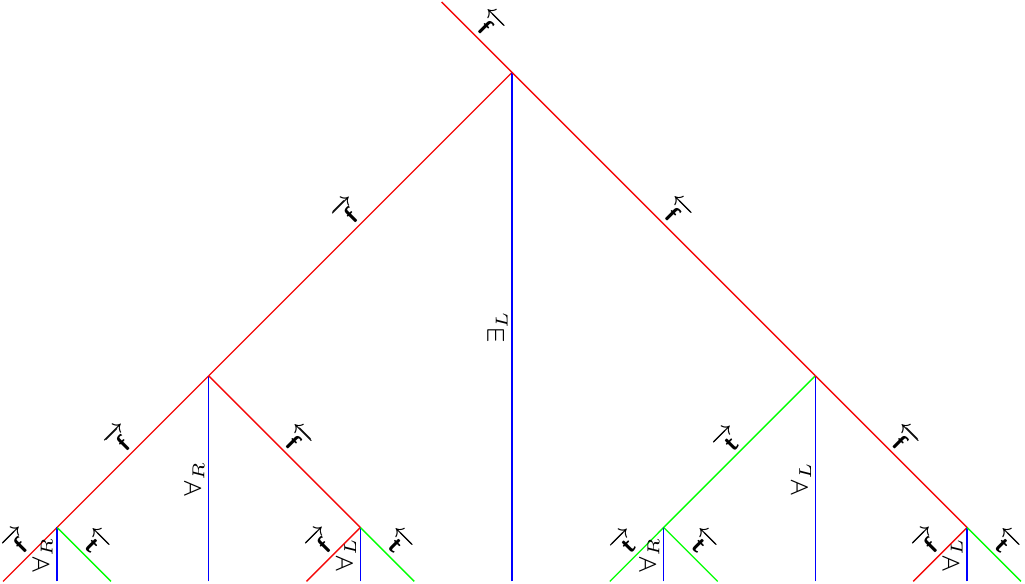}  
  \caption{Aggregation process.}
  \label{fig:app:collect}
\end{figure}

To proceed to the aggregation process, we join the results coming
from right and left two-by-two. This is done with a stationary signal indicating
the type of operation to perform --- a conjunction for $\forall$ and
a disjunction for $\exists$ --- and the direction of the resulting signal ---
left or right. The whole process is displayed in \Fig{fig:app:collect}.

\paragraph{Setting up the reduce stage:}
\[\block{reduce:qsat:init}[Q_1x_1 \cdots Q_nx_n] = \RLO{Q_n}\ldots\RLO{Q_1}\]
\begin{align*}
\ST\sigx, \LHI\sigexists & \becomes \ST\sigexistsR & 
\RHI\sigexists, \ST\sigx & \becomes \ST\sigexistsL\\
\ST\sigx, \LHI\sigforall & \becomes \ST\sigforallR &
\RHI\sigforall, \ST\sigx & \becomes \ST\sigforallL
\end{align*}

\paragraph{Executing the reduce stage:}
$\block{reduce:qsat:exec} = \RLO\sigcollect$\\

Initiation:
\begin{align*}
\RHI\sigcollect, \ST\sigt & \becomes \LLO\sigt &
\RHI\sigcollect, \ST\sigf & \becomes \LLO\sigf\\
\end{align*}

Performing the disjunction:
\begin{align*}
\RLO\sigt, \ST\sigexistsL, \LLO\sigt & \becomes \LLO\sigt &
\RLO\sigt, \ST\sigexistsL, \LLO\sigf & \becomes \LLO\sigt &
\RLO\sigf, \ST\sigexistsL, \LLO\sigt & \becomes \LLO\sigt &
\RLO\sigf, \ST\sigexistsL, \LLO\sigf & \becomes \LLO\sigf \\
\RLO\sigt, \ST\sigexistsR, \LLO\sigt & \becomes \RLO\sigt &
\RLO\sigt, \ST\sigexistsR, \LLO\sigf & \becomes \RLO\sigt &
\RLO\sigf, \ST\sigexistsR, \LLO\sigt & \becomes \RLO\sigt &
\RLO\sigf, \ST\sigexistsR, \LLO\sigf & \becomes \RLO\sigf \\
\end{align*}

Performing the conjunction:
\begin{align*}
\RLO\sigt, \ST\sigforallL, \LLO\sigt & \becomes \LLO\sigt &
\RLO\sigt, \ST\sigforallL, \LLO\sigf & \becomes \LLO\sigf &
\RLO\sigf, \ST\sigforallL, \LLO\sigt & \becomes \LLO\sigf &
\RLO\sigf, \ST\sigforallL, \LLO\sigf & \becomes \LLO\sigf \\
\RLO\sigt, \ST\sigforallR, \RLO\sigt & \becomes \RLO\sigt &
\RLO\sigt, \ST\sigforallR, \RLO\sigf & \becomes \RLO\sigf &
\RLO\sigf, \ST\sigforallR, \RLO\sigt & \becomes \RLO\sigf &
\RLO\sigf, \ST\sigforallR, \RLO\sigf & \becomes \RLO\sigf \\
\end{align*}

\begin{figure}[hbt]
  \centering
  \includegraphics[scale=.8]{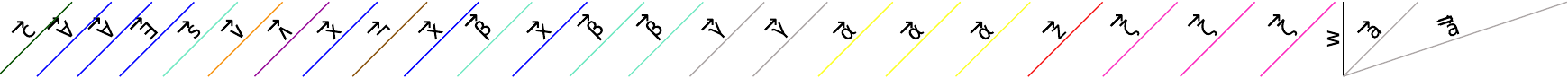}  
  \caption{Initial configuration.}
  \label{fig:app:start}
\end{figure}

Putting all the modules together, we obtain for the running example the
initial configuration shown by \Fig{fig:app:start}.

\begin{figure}[hbt]
  \centering
  \includegraphics[width=.8\textwidth]{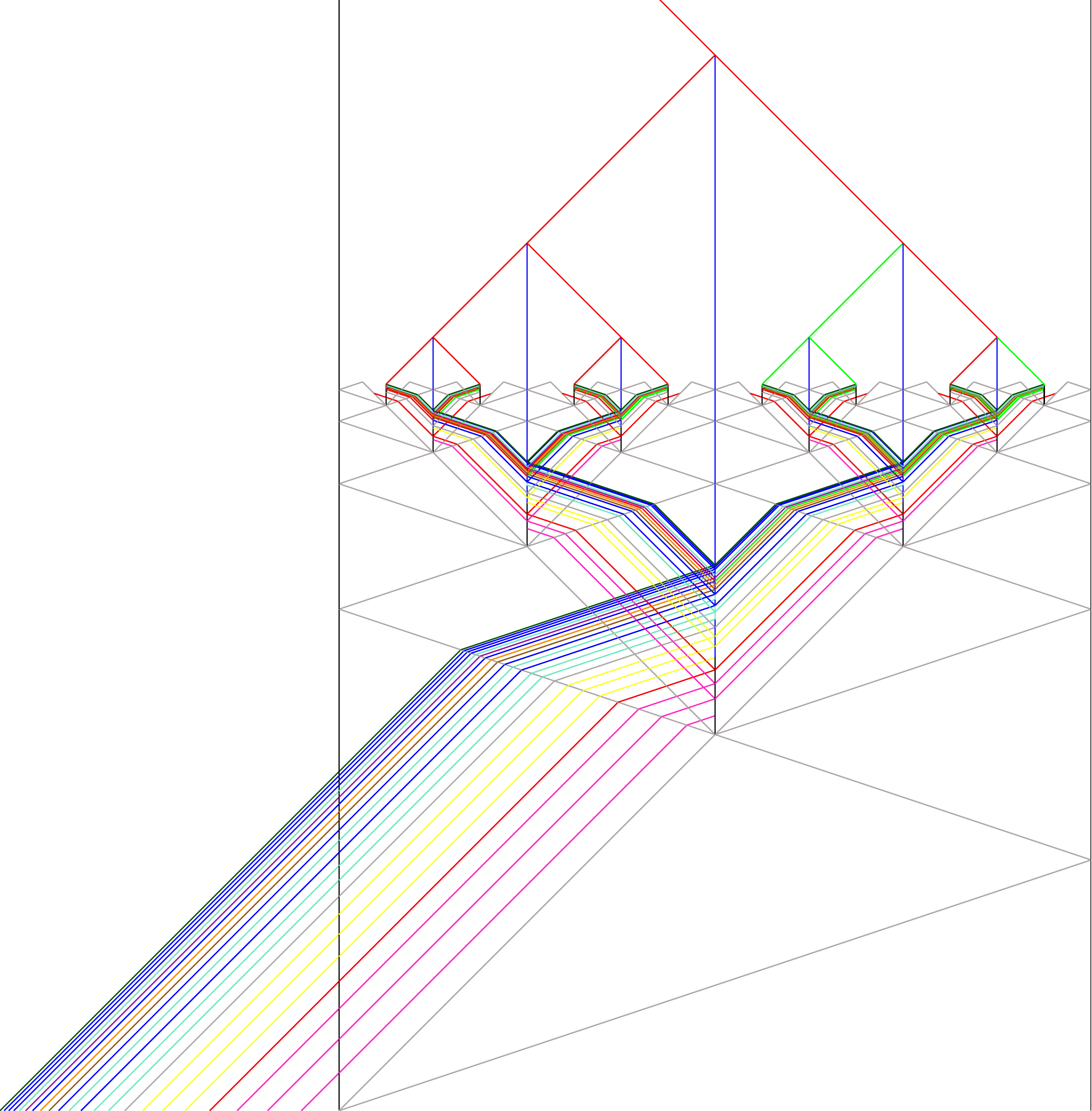}  
  \caption{The whole diagram.}
  \label{fig:app:whole}
\end{figure}

The global construction is displayed by \Fig{fig:app:whole} 

\subsection{\#SAT}

\#SAT is the problem of counting the number of
solutions of SAT. We recall that this problem is complete
for the class \#P \emph{i.e.} the class of NP-problems for which
their solutions can be counted is polynomial time.

To solve \#SAT, as a SAT-formula is a special instance of a Q-SAT formula 
in which all quantifiers are existential, we can use our Q-SAT solver and
we add a special module for counting the truth evalution of the formula
during the aggregation process. The counting is performed by a binary adder.

\paragraph{Setting up the reduce stage:}
$\block{reduce:\#sat:init} = {\RLO\sigadd}^n$
\begin{align*}
\RLO\sigadd, \ST\sigx & \becomes \ST\sigaddLZ &
\ST\sigx, \LLO\sigadd & \becomes \ST\sigaddRZ
\end{align*}

\paragraph{Executing the reduce stage:}
$\block{reduce:\#sat:exec} = \RLO\sigaddhiZ\RLO\sigaddloZ\RLO\sigzeroZ$
\begin{align*}
\RLO\sigaddloZ, \ST\sigt & \becomes \LLO\sigaddlo &
\RLO\sigaddhiZ, \ST\sigt & \becomes \LLO\sigaddhi &
\RLO\sigzeroZ, \ST\sigt & \becomes \LLO\sigone\\
\RLO\sigaddloZ, \ST\sigf & \becomes \LLO\sigaddlo &
\RLO\sigaddhiZ, \ST\sigf & \becomes \LLO\sigaddhi &
\RLO\sigzeroZ, \ST\sigf & \becomes \LLO\sigzero
\end{align*}
Rules for the binary adder (for the $R$ subscript, the rules for the $L$ are
similar, but with the output going to the left):
\begin{align*}
\RLO\sigzero, \ST\sigaddRZ, \LLO\sigzero & \becomes \ST\sigaddRZ, \RLO\sigzero &
\RLO\sigzero, \ST\sigaddRO, \LLO\sigzero & \becomes \ST\sigaddRZ, \RLO\sigone\\
\RLO\sigone, \ST\sigaddRZ, \LLO\sigzero & \becomes \ST\sigaddRZ, \RLO\sigone &
\RLO\sigone, \ST\sigaddRO, \LLO\sigzero & \becomes \ST\sigaddRO, \RLO\sigzero\\
\RLO\sigzero, \ST\sigaddRZ, \LLO\sigone & \becomes \ST\sigaddRZ, \RLO\sigone &
\RLO\sigzero, \ST\sigaddRO, \LLO\sigone & \becomes \ST\sigaddRO, \RLO\sigzero\\
\RLO\sigone, \ST\sigaddRZ, \LLO\sigone & \becomes \ST\sigaddRO, \RLO\sigzero &
\RLO\sigone, \ST\sigaddRO, \LLO\sigone & \becomes \ST\sigaddRO, \RLO\sigone\\[3mm]
\RLO\sigzero, \ST\sigaddRZ, \LLO\sigaddlo & \becomes \ST\sigaddRZ, \RLO\sigzero &
\RLO\sigzero, \ST\sigaddRO, \LLO\sigaddlo & \becomes \sigaddRZ, \RLO\sigone\\
\RLO\sigone, \ST\sigaddRZ, \LLO\sigaddlo & \becomes \ST\sigaddRZ, \RLO\sigone &
\RLO\sigone, \ST\sigaddRO, \LLO\sigaddlo & \becomes \ST\sigaddRO, \RLO\sigzero\\[3mm]
\RLO\sigzero, \ST\sigaddRZ & \becomes \ST\sigaddRZ, \RLO\sigzero &
\RLO\sigzero, \ST\sigaddRO & \becomes \ST\sigaddRZ, \RLO\sigone\\
\RLO\sigone, \ST\sigaddRZ & \becomes \ST\sigaddRZ,\RLO\sigone &
\RLO\sigone, \ST\sigaddRO & \becomes \ST\sigaddRO, \RLO\sigzero\\[3mm]
\RLO\sigaddlo, \ST\sigaddRZ, \LLO\sigaddlo & \becomes \ST\sigaddRZ, \RLO\sigaddlo &
\RLO\sigaddlo, \ST\sigaddRO, \LLO\sigaddlo & \becomes \ST\sigaddRZ, \RLO\sigone, \RHI\sigaddlo\\
\RLO\sigaddlo, \ST\sigaddRZ & \becomes \ST\sigaddRZ, \RLO\sigaddlo &
\RLO\sigaddlo, \ST\sigaddRO & \becomes \ST\sigaddRZ, \RLO\sigone, \RHI\sigaddlo\\[3mm]
\RHI\sigaddlo, \LLO\sigaddhi & \becomes \LLO\sigaddhi, \RLO\sigaddlo &
\RLO\sigaddhi, \ST\sigaddRZ & \becomes \RLO\sigaddhi
\end{align*}

\newcommand{\rsig}[1]{{\ensuremath{\overrightarrow{#1}}}\xspace}
\newcommand{\lsig}[1]{{\ensuremath{\overleftarrow{#1}}}\xspace}
\newcommand{\asig}[2]{{\ensuremath{{+}^{#1}_{#2}}}\xspace}
\begin{figure}[htb]
\centering
\begin{tikzpicture}[scale=0.4]
\draw[step=1,very thin,color=lightgray] (0,0) grid (8,15);
\draw[color=olive] (4,0) -- (4,12);
\draw (0,0) node[left] {\rsig1} -- (4,4)
      (0,4) node[left] {\rsig1} -- (4,8)
      (0,6) node[left] {\rsig A} -- (4,10)
      (0,8) node[left] {\rsig B} -- (4,12)
;
\draw (8,0) node[right] {\lsig 1} -- (4,4)
      (8,4) node[right] {\lsig A} -- (4,8)
      (8,8) node[right] {\lsig B} -- (4,12)
;
\draw[dashed,color=blue!80!black]
      (4,4) -- (7,7) node[right] {\rsig 0}
      (4,8) -- (7,11) node[right] {\rsig 0}
      (4,10) -- (7,13) node[right] {\rsig 1}
      (4.5,11.5) -- (7,14) node[right] {\rsig A}
      (4,12) -- (7,15) node[right] {\rsig B}
;
\draw[color=red!80!black]
      (4,10) -- (4.5,11.5) node[right] {\rsig C}
;
\draw[color=olive]
      (4,0) node[below] {\asig0R}
      (4,5) node[left] {\asig1R}
      (4,9) node[right] {\asig1R}
      (4,11) node[left] {\asig0R}
;
\end{tikzpicture}
\caption{Computing $3 + 1$}
\end{figure}
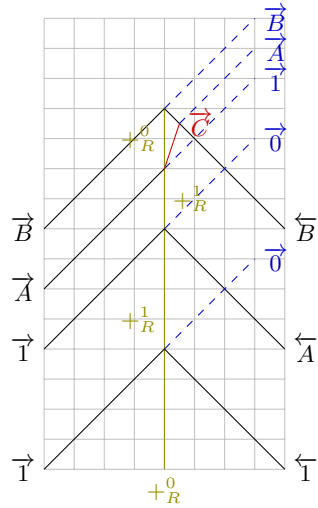

\subsection{ENUM-SAT}

ENUM-SAT is the problem of enumerating all the solutions for an instance of SAT:
we want to know \emph{all} the truth assignements of variables for which the formula is satisfiable.
We can also consider a particular case of ENUM-SAT: the problem ONESOL-SAT, which consists
in returning \emph{only one} valuation satifying the formula, when the formula is satisfiable.

\paragraph{Reduce stage:}
$\block{reduce:enumsat}[x_1 \ldots x_n] = \RLO\sigv\block{var}[x_1]\ldots\block{var}[x_n]\RLO\sigv$
\begin{align*}
\RHI\sigv, \ST\sigt & \becomes \LLO\sigv, \ST\sigv &
\RLO\sigv, \LHI\sigt & \becomes \ST\sigt, \RLO\sigv &
\RHI\sigv, \ST\sigf & \becomes \LLO\sigvZ &
\RLO\sigvZ, \LHI\sigt & \becomes \RLO\sigvZ\\
\RLO\sigv, \LHI\sigv & \becomes \ST\sigv &
\RLO\sigv, \LHI\sigf & \becomes \ST\sigf, \RLO\sigv &
\RLO\sigvZ, \LHI\sigv & \becomes &
\RLO\sigvZ, \LHI\sigf & \becomes \RLO\sigvZ
\end{align*}

\subsection{MAX-SAT}
The problem MAX-SAT consists in, given $k$ SAT-formulae, finding the maximum
number of formulae satisfiable by the same valuation.  This problem is NP-hard
and is complete for the class APX --- the class of problems approximable in
polynomial time with a constant factor of approximation.  The problem MAX-SAT
can be extended by returning the valuation of variables that satisfies the
greater number of formulae amoung the $k$ ones.

We do not give the corresponding rules for solving MAX-SAT, we just describe 
the concerned modules. Each formula amoung the $k$ formulae given in the input
is compiled by the same method used previously, and the resulting 
arrangement of signals for each formula are placed end-to-end. This
results in a beam of formulae composed by $k$ sub-beam, one for each formula.
The evaluation process is then the same as seen previously.

To compare the number of satisfiable formulae for each valuation, we used
the binary adder introduced for \#SAT, that we combine with a module comparing
two binary number. The reduce phase follows the same idea that for the other variants, 
except that after comparing two-by-two the number of formulae satisfiable, 
the greater number is transmitted to the next level of agregation for the next comparison.
Then, at the top of the construction, we can read in the  binary representation
of the maximal number of satisfiable formulae.

If we also want the truth assignement that satisfies the greater number of
formulae, we can easily devise a new module on the basis of the one used for
ENUM-SAT.
%


\end{document}